\pdfoutput=1
\documentclass[12pt,pdftex]{article}
\usepackage[margin=0.5in]{geometry}
\usepackage{color}
\usepackage[pdftex]{graphicx,graphics}
\usepackage{authblk}
\title{Extended use of superconducting magnets for bio-medical development}
\author{Stoyan Stoynev\thanks{stoyan@fnal.gov}}
\affil{FNAL}

\begin{document}
\maketitle

\abstract{
Magnetic fields interact with biological cells affecting them  in variety of ways which are usually hard to predict.
Among them, it was observed that strong fields can align dividing cells in a preferred
direction. It was also demonstrated that dividing cancer cells are effectively destroyed by  
applying electric fields in vivo with a success rate dependent on the cell-to-field orientation. Based on these facts, 
the present note aims to suggest the use of 
magnetic and electric fields for improved cancer treatment. Several possibilities of generating the electric fields 
inside the magnetic field volume are reviewed, 
main tentative approaches are described and discussed. Most if not all of them require special magnet configuration 
research which can be based on existing magnet systems in operation or in development.
}


\section{Introduction}

Developments in High Energy Physics (HEP) have significant impact on 
our society though it may not be well realized. 
For understandable reasons, the most recognizable and appreciable contributions are arguably in the field of medicine. 
Particle beams are used for cancer treatment, isotope production and equipment sterilization ; particle detectors are 
used for imaging diagnostics (PET scanners) and strong superconducting magnets (typically researched and then produced ``en masse'' for 
HEP accelerators) are used in magnetic resonance imaging (MRI) systems. The significance of these applications and the 
fact that they, inevitably, rely on interdisciplinary knowledge and support of sciences can not be overstated.
Such an interaction is healthy for the respected fields and is probably a key for faster progress in many areas of life 
and science. Not least it is much better accepted in view of the public opinion than pure field applications which is important 
for the support of fundamental studies.

One of the mentioned contributions of HEP is the development of magnets of variable complexity. Such devices are part of 
MRI which guiding principle is basically based on the re-alignment of (hydrogen) nuclei in a magnetic field.
This effect is well founded and precisely calculable. It is much more difficult to predict and even more to calculate what 
the bio-physical effect of magnetic field on living organisms would be. Experiments conducted on living organisms in fact show that at least some 
form of bio-cell alignment is observed in many cases.  These include but are not limited to red blood cells \cite{bloodMagA}, smooth 
muscle cells \cite{muscleMagA}, nerve cells \cite{nerveMagA}, bone (collagen, osteroblasts) \cite{boneMagA, boneMagA2}, bio-crystals 
(in fish and algae) \cite{fishMagA}, 
frog eggs \cite{frogMagA}, yeast \cite{yeastMagA} and even particular cancer cells studied \cite{cancerMagA}.    
The studies suggest that stronger the field strength 
better the (average) cell alignment with it. This is not surprising by itself as any (diamagnetic) material but also ion flow would react to 
a sufficiently strong magnetic field. Symmetries in cells are what makes the process worth the attention in this note.  Quickly dividing cell cultures, like yeast, 
are of 
particular interest as dividing cells by definition posses clearly defined symmetry axes. 
The fact that such cells align 
is easier to observe, explore and apply to cell populations sharing similar features. Analogy to cancer cells is implied though  
ultimately every cell culture of interest needs to be tested. 

Fairly recently it was shown that alternating electric fields, above some intensity and with optimal frequency, fight 
cancer \cite{ttf}. It was also emphasized that the effects (albeit with hard to strictly prove underlining mechanisms) of cell 
proliferation and cell growth rates were dependent on the cell alignment to the electric field. 
It is clear that if cell orientation could have been enforced results from electric field exposure would have been improved as well. Magnetic fields can be used as such an alignment tool. However they are much more than that - they themselves 
can induce the electric potential needed.
Magnetic fields present the opportunity to enhance the applications and extend of cancer treatment by exploring the 
intrinsic physical duality of the electro-magnetic field. 
   
It is fair to shed some light on the interaction between magnetic fields with living organisms. 
It is generally agreed there are no lasting effects of fields up to at least 8 T on individual cells, 
cell structures or metabolism (see for instance US Food and Drug Administration /FDA/ Guidelines in \cite{fda}). 
However variety of significant temporary effects do exist and in some cases cell cultures are 
severely affected. Effects are discussed in ref.
\cite{review_magCan}, \cite{studies_magBio}, \cite{review_cellMag1}, \cite{review_cellMag2}, \cite{survive_varMag}, \cite{study_13T}.
Complete systematic studies on the effects of magnetic fields (strong dipole, gradient, variable/pulsed) on living tissue are in fact not 
available. As the reviews referenced also suggest, magnetic fields by their own affect cancer growth though studies are incomplete and 
inconclusive. 
It is worth emphasizing the value of such 
investigations - they are a likely by-product of the type of studies suggested by this document. 

The current note discusses particular applications of strong external magnetic fields, in combination or not with electric and 
acoustic fields, to quickly expanding (dividing) bio-tissue. It is argued that they greatly enhance the available approaches to fight 
cancer given already proved mechanisms of anti-cancer action. It is also argued that a successful development requires technological 
research on magnets as well as extended collaborations with expertise in variety of fields in science.

\section{Targeting the process of cell division}

One of the main discriminators between cancer and non-cancer cells is the continuous rapid 
reproduction of the former and thus the fact that cancer cells are much more often in a state of division.
Targeting cell/nucleus division (mitosis) and the structures supporting it was recognized as an anti-cancer tool 
long ago (see \cite{targetSpindle}) and various approaches along this line were explored. More
recently it was suggested that the targeting can be accomplished by alternating electric fields \cite{ttf}.
Indeed, the method (a.k.a TTF or Tumor Treating Field) is now accepted in the USA by the FDA
and in other parts of the world. The suggested physics explanation behind the success of this new 
treatment has to do with the inhomogeneity of the induced electric field inside a cell 
during mitosis together with the natural alignment of highly polarized molecules 
and likely the ion flow in the process. As a result of applied field of  $\sim$ 1 V/cm, with a chosen frequency 
of hundreds of kHz, extended period for the mitosis and ultimately cell death were observed.
As it was emphasized the field achieved slight alignment of the cells but overall multiple sources 
from different directions had to be applied to mitigate the effect of random angle distribution. As demonstrated 
and as expected cells aligned to the direction of the electric field were most susceptible to destruction.

\begin{figure}
{\centering
\includegraphics[width=0.48\textwidth]{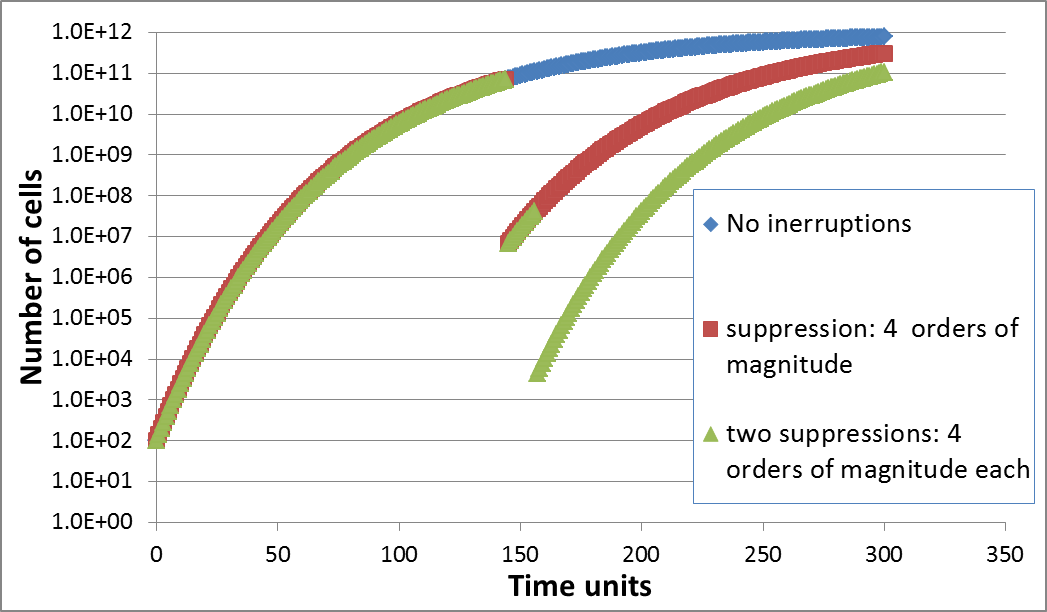}
\caption{\label{growth}
Uncontrolled cell growth according to the Gompertz model (see text) with parameters  
$N_{\infty} = 10^{12}$, $B=10$ and 
$C = 0.015  (time~unit)^{-1}$. Two instantaneous suppression models are shown together with no suppression 
model. The figure is for illustrative purposes only. 
}
}
\end{figure}

The TTF method shows (overall) the same success rate as drugs (chemotherapy)\footnote{However there are much less adverse 
effects with TTF and thus the quality of life is better.}
 and in combination they are shown to improve the medical outcomes \cite{ttf2}. 
Most treatments, even if very successful,  could not continue indefinitely for variety of reasons. 
It is worth emphasizing a general feature of tumor masses that makes them hard to deal with. 
Starting with relatively 
small number of cells the number grows exponentially but as cells have to compete for resources 
(like limited number of blood cells feeding them) a limit is slowly reached for that cell culture. Although there is no universal 
description for the process the Gompertz model can be used to approximate it \cite{Gompertz}:

\begin{equation}
N(t) = N_{\infty}e^{-Be^{-Ct}}
\end{equation}

, where $N_{\infty}$ is the asymptotic limit, $B$ defines the initial condition, $C$ is the growth rate and $t$ is time.
Figure \ref{growth}
compares the cell growth development for unperturbed population and two cases with one and two 
instantaneous population reductions of 99.99\% (each is a ``4-log-kill'' which would be considered a very very 
successful outcome). 
Not only the population quickly recovers but, as 
the growth rate depends on the population, higher the kill rate quicker the recovery rate!
Such a figure suggests why only a complete cancer cell obliteration is a long term remedy.

\section{Magnetic field applications}

There are studies showing that the magnetic field - constant, gradient or alternating (or pulsed) - itself 
could have anti-cancerous effect. A review can be found it ref. \cite{review_magCan}. The current note is not 
about these but rather about the simultaneous 
application of external magnetic field for proper cell alignment and an electrical field, external or not, 
on the sensitive cell ingredients. As mentioned, in first order even very strong magnetic fields are harmless 
to multi-cell living organisms and field strengths of at least several tesla start to impose significant 
alignment on dividing cells depending on their type. Then considering the presence of magnetic field we find 
additional 
ways to induce electric field on or inside the dividing cells and thus disrupt the process of division. Hence 
the set of available 
options is enriched and the probability that technological obstacles can be overcome is increased. 

There are more subtle points about (dividing) cell orientation that are best to be understood when a 
technique is developed.
Even though prefered orientation is statistically confirmed in cultures we ultimately need all cells to be 
properly aligned. In order to do that an understanding of the process for individual cells
is necessary. 
As observed in many cases a cell division plane follows symmetry laws - the cell 
orientation in absence of external forces depends on the cell orientation of the previous generation(s). 
Ref. \cite{magOrientModel} presents a mathematical 
model of cell division orientation from generation to generation for a particular culture (frog eggs)
with or without influence of magnetic field.  Cell orientation is modeled following observations.
%
A more detailed practical method for mitotic spindle orientation analysis is suggested in ref. 
\cite{modelSpindleOrient}.
It is important to find how much 
in common different relevant cultures have and more importantly to try to understand where differences from 
physical point of view come from. 
Models describing the effects of magnetic fields and possibly understanding 
of the processes would allow for much better 
planning of the type and strenghts of 
magnetic fields needed to exert an optimal alignment effect on the cultures in question. Throughout the text we 
will neglect these fine but important details 
and just assume that the magnetic fields orient cells with some (high) efficiency.    

The electric field applied in the TTF method acts best when dividing cells have their division plane perpendicular to 
the electric field direction. However it was also shown that there were significant effects on cells 
even when the electric field was parallel to the division plane. This brings up two important points. Ideally a 
magnetic field aligns dividing cells such that the division plane is perpendicular to the field direction.
Then an electric field would be applied that is perpendicular to the division plane (so parallel to the 
magnetic field in this case). Both cases are not necessarily unique in all cultures. Although there is 
seemingly cylindrical symmetry in a dividing cell internal structures may have hidden symmetries. 
Still it is feasible to assume that the division plane is a symmetry plane. Hence in general 
a magnetic field is expected to orient cell division planes either perpendicular or parallel (due to internal 
symmetries) to the magnetic field. Then the electric field disrupts the process of division very efficiently if
normal to the division plane, less efficiently when parallel to it and the effect is further 
suppressed when in mid-angle situation \cite{ttf}. Combining expectations and observations, we can conclude 
that for a given culture the requirement is to have the electric and the aligning magnetic field either parallel 
or perpendicular to each other, with the former most often the case.

%
  
\subsection{External magnetic and electric fields}

An obvious though potentially technically challenging approach is the separate application of a 
strong constant dipole magnetic field 
and variable electric field. Ideally the only difference with TTF is the presence of (properly 
oriented) magnetic field.
It maybe be difficult to operate a finely tuned electrical device inside a strong magnetic 
field 
If this is the case  an option is to 
apply the magnetic and then the electric 
field consequently. It is applicable 
as long as the characteristic times of de-orientation of bio-molecules is much larger that the periods 
of time with no magnetic field. Typical times of de-orientation of cells are high, often hours \cite{cellTimeOrient}.
De-orientation of the mitotic apparatus, if misaligned with respect to natural markers, can takes 
ten and more minutes \cite{spindleTimeOrient}.
Thus the approach with variable alignment field should be applicable for cycles of at least several minutes 
and possibly much more. 
With or without interruptions in the magnetic field the method is 
the most straightforward to implement
and can be a proof of concept as long as it is rendered technically operational. A drawback of the method (just as for TTF) is 
that the area where the electric field is applied can not be customized easily and usually includes undesired regions 
of space.

\subsection{External magnetic and acoustic fields}

It was already suggested that magnetic and acoustic fields can be used together for cancer treatment \cite{magSound} though 
cell alignment to the magnetic field was not considered. Acoustic pressure causes molecular movements in tissue and in the 
presence of magnetic field which is not parallel to the movement direction an
electric field perpendicular to both the movement and magnetic field directions is induced. 
The following equations relate the acoustic pressure $p$ with particle velocity $v_{part}$ in a medium and 
the intensity $I$ of the ultrasound field:

\begin{equation}
p = Z v_{part}
\end{equation}

\begin{equation}
I = pv_{part}
\end{equation}

with $Z$ the specific acoustic impedance of the medium. For water (and approximately for tissue) $Z = 1.5 MN s/m^3$. At $I = 100 W/cm^2$ 
(the most powerful pulsed devices achieve two orders of magnitude more \cite{ultraSoundIntensity}) one gets 
$p \sim 1 MPa$ and $v_{part} \sim 1 m/s$.
We associate that velocity to the amplitude of the cell velocity during the cell's cyclic movement
under the sound pressure.
The following equation gives the induced electric field $E$ 
in the cell 
in the presence of magnetic field with flux density $B$:

\begin{equation}
E = B v_{part}
\end{equation}

In 10 T field the maximal electric field amplitude is $E = 0.1 V/cm$. Although it is lower that the minimal 
external 
electric field giving the desired effects in TTF it can be regulated by the acoustic intensity and also it should be noted that lower 
induced electrical fields may be needed to achieve the same results with respect to external fields (where the membrane capacitance plays an important role). This subject 
 can be addressed only experimentally. The point being made is that 
experimental approaches to resolve the problem are within reach. 

To minimize the effect of absorption the ultrasound should be with lower frequency. Even few tens of kHz allows for 
a fairly well spacial resolution. 
To maximize the effect in the desired region and suppress it elsewhere multiple (properly synchronized) acoustic 
sources could be used and directed toward the area of interest (as noted in \cite{magSound} as well). To minimize 
undesired effects of sound reflection tissue and the frond-end of the acoustic device(s) should have similar 
specific acoustic impedance with air excluded as a medium. One of the advantages of this method is that a small,
of the order of one cubic cm, region of space can be targeted by the ultrasound with either crossing collimated beams 
or a focused beam, leaving the remaining area less affected.         

The estimations performed above imply that the magnetic and the induced electric fields are perpendicular to each other.
There is clearly an obstacle in the case where the aligning magnetic field and the induced electric field should
be in parallel. To resolve this problem we consider a constant aligning dipole magnetic field and a variable 
(or pulsed) dipole magnetic field in perpendicular direction (that is - a ``skew'' dipole field component). Then 
each magnetic and the induced electric fields are perpendicular to each other with the sound direction 
in parallel to the alignment field. The variable/pulsed magnetic field should be strong enough (order of 1-10 T) 
and still devised in such a way as to leave the average cell orientation induced by the alignment field 
unperturbed \footnote{As mentioned, this can be achieved because the processes associated with cell redistribution 
are relatively slow}. Nevertheless when multiple sound sources are used both magnetic fields have contributions 
to the 
induced electric field inside the cell creating a more complex distribution of it (with unknown consequences
to the cell division process).

\subsection{Variable magnetic fields}

We consider a time-dependent dipole field in the direction of the cell division axis - the case with perpendicular 
field (``skew'' component) was already discussed. 
Variable magnetic field induces (non-conservative) electric field in the bio-tissue. This variation could be a 
modulation 
on top of a strong dipole field and can consists of large amplitude pulses. Modulation of the field will likely 
mean two separate magnets otherwise 
technological difficulties will prevail. 
To give an estimate of the induced electric field we assume that a dividing cell has a characteristic length
(diameter) $d$ and that it defines the relevant sensitive area. The electric field $E$ induced at the cell 
is given by 
\begin{equation}
E \sim \frac{d}{4} \frac{\partial B}{\partial t} 
\end{equation}
and if we assume that 1 V/cm is the desired magnitude for the induced electric filed 
then ~0.1 T/ms rate of change of the field is needed to reach it. This is very much achievable in 
pulsed magnets where values of tens of tesla are reached within tens of milliseconds \footnote{There are many pulsed magnet facilities 
in the world; one of the leading ones is the Los Alamos National Laboratory in New Mexico - www.lanl.gov}. These are one or two orders of 
magnitude higher that needed for the purposes discussed. 
The combined effect 
of the pulsed magnetic field and the constant field should be strong enough to align cells which gives 
additional degrees of freedom in terms of relative field strenghts and variable field frequency and duty cycle.
An alternative to this approach is to apply two variable uni-directional magnetic fields - one pulsed with low 
frequency and another operating during the ``non-active'' part of the cycle of the first magnet. Such a
solution may be preferable with respect to simultaneous operation of the two magnets.       


\subsection{Gradient and higher order harmonics magnetic fields}

Electrical field will be induced by gradient magnetic field in moving bio-cells based on the following relation:
$E \sim \frac{d}{4} \frac{\partial B}{\partial s} v_{part}$ , where $\frac{\partial B}{\partial s}$ 
is the field gradient.
Typical velocities and achievable field gradients result in orders of magnitude lower electric fields 
compared to the methods already discussed. Gradient fields could exert additional force on molecules due to their
magnetic dipole moments and if any, this is the effect that might be of use in obstructing cell division. 
In practice large 
gradients can be created by multipole magnets.
Higher order field harmonics themselves are of no direct gain except if very peculiar field 
configurations are required.  
We are not considering these here.       

\subsection{Technical aspects about magnets}

In the most trivial case a strong dipole magnet with sufficiently large 
aperture should be able to operate for hours. The magnetic field should be able to 
reach at least 10 T so that field strength dependent effects could be investigated - this also 
concerns all other cases later.
A high frequency electrical TTF-type device should be able to operate inside the field.
There is nothing special about the magnet requirements and as long as the TTF device
is operational the configuration can be build by the help of existing magnets or 
magnet prototypes. In most cases here and later we are implying superconducting devices though 
there may as well be exceptions (in particular - pulsed magnets). 

The first development issue comes with the requirement to have two perpendicular field 
components with one of them dominant (it should define the cell alignment axis). These fields 
should be delivered to the same area simultaneously or semi-simultaneously (relying on short period cell 
immobility). This is not a regular requirement and magnets of such a type should be manufactured. 
Moreover the second magnet should be either with variable magnetic field or based on high power 
pulses. Especially in the latter case this is likely to affect the normal operation of the first 
superconducting magnet. The most probable solution is to operate the two magnets in repeated 
cycles (thus both of them will be producing variable fields). There are no obvious obstacles 
 to mechanically devise 
the two magnets fitting together as the two fields are perpendicular to each other. 
The more challenging task is to resolve potential operational issues.

If we only need uni-directional two-component field this still means two separate magnets
as high frequency/pulsed magnets come with much different parameters than constant field magnets 
(notably - inductance). In effect we search for such a configuration as to produce a modulated 
magnetic field in the area of interest. This is not a trivial task given the desired characteristics of the 
field - essentially a strong (semi-)constant component and powerful pulses with the same field direction. 
It is a major technological challenge to overcome but there is nothing suggesting it can not be done.

The above gives the basic idea of the type of developments needed in order to make further progress.
The technological means are available and most often the requirements are away from the  
edge of technology. To fully appreciate 
future experiment outcomes we argue that the complete set of options should be considered and tested.      

\section{Research and collaboration}

Because of the strong magnetic field required only superconducting devices can be used for tests.
Superconducting magnets are typically expensive and not trivial to manufacture and operate. On 
the other hand existing devices, including very high field MRI and many HEP magnets, would meet the 
loose requirements for initial tests. As discussed a systematic study requires different types of devices 
possibly with characteristics to be varied. 
The optimal approach to this problem is to use facilities that have good expertise and practise with such devices and
eventually operate/test variety of them. As already mentioned HEP experiments require huge production 
of superconducting magnets with different characteristics. It seems a viable collaboration between 
HEP organizations involved in magnet fabrication  on one side and biology/medical 
institutions on another can be formed. If properly planned such a collaboration will inevitably deliver 
valuable results and hopefully crucial observations and insights for further development.     

Effects of magnetic field on living organisms and on humans in particular are not fully understood
and there is no clarity on the safety of strong (including gradient/alternating) magnetic fields. There is 
a lack of systematic investigation and safety rules guidance does not extrapolate far enough. 
Yet stronger and stronger fields are tested and some of them are also needed for medical purposes.
The point of systematic information gathering, likely needed for performing the studies outlined and thus
constituting a by-product, has its own significance. It could easily be a part of a more general systematic 
research on living organisms in strong magnetic fields though this is not necessarily related to the 
main subject of this note.
It is related however to the mentioned type of collaborations needed to accomplish such a goal.

\section{Conclusions}

We have presented and discussed the possibility of more efficient cancer treatment based on 
external magnetic, electric 
and acoustic fields. All revolve around the ability of strong magnetic fields to align dividing cells 
combined with  simultaneous application of electric field to disrupt the cell division. Both effects are known 
in the literature and in practice. A sufficiently strong 
magnetic fields are needed which requires superconducting devices and supporting facilities. 
Several viable options exist to create the necessary electric field and part of them necessitate 
development of magnets with special characteristics. It was shown that overcoming technical obstacles 
is withing the reach of current technologies. A comprehensive research and analysis 
implies forming strong interdisciplinary collaborations.

\section*{Acknowledgments}

Operated by Fermi Research Alliance, LLC under Contract No. DE-AC02-07CH11359 with the United States Department of Energy.

I'd like to thank George Velev and Sandor Feher for useful discussions. 

%
%
%
%
%
%

\bibliographystyle{unsrt}
\bibliography{magnetCan}

\begin{thebibliography}{10}

\bibitem{bloodMagA}
T.~Higashi, A.~Yamagishi, T.~Takeuchi, N.~Kawaguchi, S.~Sagawa, S.~Onishi, and
  M.~Date.
\newblock {Orientation of erythrocytes in a strong static magnetic field}.
\newblock {\em Blood}, 82(4):1328--1334, 1993.

\bibitem{muscleMagA}
Masakazu Iwasaka, Junji Miyakoshi, and Shoogo Ueno.
\newblock {Magnetic field effects on assembly pattern of smooth muscle cells}.
\newblock {\em In Vitro Cell. Dev. Biol.}, 39(3-4):120--123, 2003.
\newblock doi:10.1007/s11626-003-0005-0.

\bibitem{nerveMagA}
Yawara Eguchi, Mari Ogiue-Ikeda, and Shoogo Ueno.
\newblock {Control of orientation of rat Schwann cells using an 8-T static
  magnetic field}.
\newblock {\em Neurosci. Lett.}, 351(2):130--132, 2003.
\newblock doi:10.1016/S0304-3940(03)00719-5.

\bibitem{boneMagA}
Hiroko Kotani, Hiroshi Kawaguchi, Takashi Shimoaka, Masakazu Iwasaka, Shoogo
  Ueno, Hidehiro Ozawa, Kozo Nakamura, and Kazuto Hoshi.
\newblock {Strong Static Magnetic Field Stimulates Bone Formation to a Definite
  Orientation In Vitro and In Vivo}.
\newblock {\em J. Bone Miner. Res.}, 17(10):1814--1821, 2002.
\newblock doi:10.1359/jbmr.2002.17.10.1814.

\bibitem{boneMagA2}
J.~Torbet and M.C. Ronziere.
\newblock {Magnetic alignment of collagen during self-assembly}.
\newblock {\em Biochem J.}, 219(3):1057–1059, 1984.

\bibitem{fishMagA}
Yuri Mizukawa and Masakazu Iwasaka.
\newblock {Magnetic Orientational Tweezers for Cell Manipulation}.
\newblock {\em Transactions of Japanese Society for Medical and Biological
  Engineering}, 51(Supplement):M--127, 2013.
\newblock doi:10.11239/jsmbe.51.M-127.

\bibitem{frogMagA}
James~M. Denegre and James~M. Valles.
\newblock {Cleavage planes in frog eggs are altered by strong magnetic fields}.
\newblock {\em Proc. Natl. Acad. Sci. USA}, 95:14729--14732, 1998.

\bibitem{yeastMagA}
Shigeki Egami, Yujiro Naruse, and Hitoshi Watarai.
\newblock {Effect of static magnetic fields on the budding of yeast cells}.
\newblock {\em Bioelectromagnetics}, 31(8):622--629, 2010.
\newblock doi:10.1002/bem.20599.

\bibitem{cancerMagA}
Laura Teodori, Maria~C. Albertini, Francesco Uguccioni, Elisabetta Falcieri,
  Marco B.~L. Rocchi, Michela Battistelli, Carlo Coluzza, Giovanna Piantanida,
  Antonio Bergamaschi, Andrea Magrini, Raffaele Mucciato, and Augusto Accorsi.
\newblock {Static magnetic fields affect cell size, shape, orientation, and
  membrane surface of human glioblastoma cells, as demonstrated by electron,
  optic, and atomic force microscopy}.
\newblock {\em Cytometry Part A}, 69A(2):75--85, 2006.
\newblock doi:10.1002/cyto.a.20208.

\bibitem{ttf}
Eilon~D. Kirson, Zoya Gurvich, Rosa Schneiderman, Erez Dekel, Aviran Itzhaki,
  Yoram Wasserman, Rachel Schatzberger, and Yoram Palti.
\newblock {Disruption of cancer cell replication by alternating electric
  fields}.
\newblock {\em Cancer Res.}, 64(9):3288--3295, 2004.
\newblock doi:10.1158/0008-5472.CAN-04-0083.

\bibitem{fda}
U.S. Food and Drug Administration.
\newblock Criteria for significant risk investigations of magnetic resonance
  diagnostic devices - guidance for industry and food and drug administration
  staff, 2014.

\bibitem{review_magCan}
Soumaya Ghodbane, Aida Lahbib, Mohsen Sakly, and Hafedh Abdelmelek.
\newblock {Bioeffects of Static Magnetic Fields: Oxidative Stress, Genotoxic
  Effects, and Cancer Studies}.
\newblock {\em BioMed Research International}, 2013:602987, 2013.
\newblock doi:10.1155/2013/602987.

\bibitem{studies_magBio}
D.~Rosen.
\newblock {Studies on the Effect of Static Magnetic Fields on Biological
  Systems}.
\newblock {\em PIERS Online}, 6(2):133--136, 2010.
\newblock doi:10.2529/PIERS090529114533.

\bibitem{review_cellMag1}
Junji Miyakoshi.
\newblock {The review of cellular effects of a static magnetic field}.
\newblock {\em Sci. Technol. Adv. Mater.}, 7:305--307, 2006.
\newblock doi:10.1016/j.stam.2006.01.004.

\bibitem{review_cellMag2}
Junji Miyakoshi.
\newblock {Effects of static magnetic fields at the cellular level }.
\newblock {\em Prog. Biophys. Mol. Bio.}, 87(2-3):213–223, 2005.
\newblock doi:10.1016/j.pbiomolbio.2004.08.008.

\bibitem{survive_varMag}
J.~Lipiec, P.~Janas, and W.~Barabasz.
\newblock {Effect of oscillating magnetic field pulses on the survival of
  selected microorganisms}.
\newblock {\em Int. Agrophysics}, 18(4):325--328, 2004.

\bibitem{study_13T}
Zhao Guoping, Chen Shaopeng, Zhao Ye, Zhu Lingyan, Huang Pei, Bao Lingzhi, Wang
  Jun, Wang Lei, Wu~Lijun, Wu~Yuejin, and Xu~An.
\newblock {Effects of 13 T Static Magnetic Fields (SMF) in the Cell Cycle
  Distribution and Cell Viability in Immortalized Hamster Cells and Human
  Primary Fibroblasts Cells}.
\newblock {\em Plasma Science and Technology}, 12(1):123, 2010.
\newblock doi:10.1088/1009-0630/12/1/26.

\bibitem{targetSpindle}
K.~W. Wood, W.~D. Cornwell, and J.~R. Jackson.
\newblock {Past and future of the mitotic spindle as an oncology target}.
\newblock {\em Current Opinion in Pharmacology}, 1(4):370--377, 2001.
\newblock doi:10.1016/S1471-4892(01)00064-9.

\bibitem{ttf2}
Eilon~D. Kirson, Rosa~S. Schneiderman, Vladimír Dbalý, František Tovaryš,
  Josef Vymazal, Aviran Itzhaki, Daniel Mordechovich, Zoya Gurvich, Esther
  Shmueli, Dorit Goldsher, Yoram Wasserman, and Yoram Palti.
\newblock {Chemotherapeutic treatment efficacy and sensitivity are increased by
  adjuvant alternating electric fields (TTFields)}.
\newblock {\em BioMed Central Med. Phys.}, 9(1), 2009.
\newblock doi:10.1186/1756-6649-9-1.

\bibitem{Gompertz}
Charles~P. Winsor.
\newblock {The Gompertz Curve as a Growth Curve}.
\newblock {\em Proc. Natl. Acad. Sci. U S A}, 18(1):1--8, 1932.

\bibitem{magOrientModel}
James~M. Valles~Jr.
\newblock {Model of magnetic field-induced mitotic apparatus reorientation in
  frog eggs}.
\newblock {\em Biophys J.}, 82(3):1260–1265, 2002.

\bibitem{modelSpindleOrient}
Christoph Jüschke, Yunli Xie, Maria~Pia Postiglione, and Juergen~A. Knoblich.
\newblock {Analysis and modeling of mitotic spindle orientations in three
  dimensions}.
\newblock {\em Proc. Natl. Acad. Sci. U.S.A.}, 111(3):1014--1019, 2014.

\bibitem{cellTimeOrient}
R.~Kemkemer, C.~Neidlinger-Wilke, L.~Claes, and H.~Gruler.
\newblock {Cell orientation induced by extracellular signals}.
\newblock {\em Cell Biochem. Biophys.}, 30(2):167--192, 1999.

\bibitem{spindleTimeOrient}
Mickael Machicoane, Cristina A.~de Frutos, Jenny Fink, Murielle Rocancourt,
  Yannis Lombardi, Sonia Garel, Matthieu Piel, and Arnaud Echard.
\newblock {SLK-dependent activation of ERMs controls LGN–NuMA localization
  and spindle orientation}.
\newblock {\em J. Cell Biol.}, 205(6):791–799, 2014.
\newblock doi:10.1083/jcb.201401049.

\bibitem{magSound}
Friedwardt Winterberg.
\newblock {Treating Cancer with Strong Magnetic Fields and Ultrasound}.
\newblock arXiv:0906.0742 [physics.gen-ph], 2009.

\bibitem{ultraSoundIntensity}
Michael~S. Canney, Michael~R. Bailey, and Lawrence~A. Crum.
\newblock Acoustic characterization of high intensity focused ultrasound
  fields: A combined measurement and modeling approach.
\newblock {\em J. Acoust. Soc. Am.}, 124(4):2406–2420, 2008.
\newblock doi:10.1121/1.2967836.

\end{thebibliography}
\end{document}